%% file: ms.tex
\newif\ifapj    
\apjfalse

\newif\ifpdflatex    
\pdflatextrue		

\ifapj
  \documentclass[apj]{emulateapj}
  \usepackage{graphicx}
  \usepackage{epsfig}
  \usepackage{portland}
\else
  \documentclass[usenatbib]{mn2e}
  \usepackage{graphicx}
  \def\lesssim{\mathrel{\hbox{\rlap{\hbox{\lower5pt\hbox{$\sim$}}}\hbox{$<$}}}}
  \def\gtrsim{\mathrel{\hbox{\rlap{\hbox{\lower5pt\hbox{$\sim$}}}\hbox{$>$}}}}
\fi

\usepackage[normalem]{ulem}
\newcommand{\mytitle}{LOSS's First Supernova? New Limits on the ``Impostor" SN 1997bs}
\newcommand{\myshorttitle}{Imposter SN 1997bs}
\newcommand{\myshortauthors}{Adams \& Kochanek}

\ifapj
  \shortauthors{\myshortauthors}
  \shorttitle{\myshorttitle}
  
\begin{document}
  \title{\mytitle}
  \author{Scott M. Adams\altaffilmark{1} \& C.S. Kochanek\altaffilmark{1,2}}
  \altaffiltext{1}{Department of Astronomy, The Ohio State University, 140 W.\ 18th   Ave., Columbus, OH 43210, USA}
  \altaffiltext{2}{Center for Cosmology and AstroParticle Physics (CCAPP), The Ohio State University, 191 W.\ Woodruff Ave., Columbus, OH 43210, USA}
\else
  \title[\myshorttitle]{\mytitle}
  \author[\myshortauthors]
	{\parbox{18cm}{Scott M. Adams$^{1}$ \& C.S. Kochanek$^{1,2}$}
  \\
  \\
  $^{1}$ Deptartment of Astronomy, The Ohio State University, 140 W.\ 18th   Ave., Columbus, OH 43210, USA\\
  $^{2}$ Center for Cosmology and AstroParticle Physics (CCAPP), The Ohio State University, 191 W.\ Woodruff Ave., Columbus, OH 43210, USA\\
  E-mail: sadams@astronomy.ohio-state.edu}
  \begin{document}
  \voffset -1.5cm
  \maketitle
\fi

\begin{abstract}
We present new, late-time \emph{Hubble Space Telescope} and \emph{Spitzer Space Telescope} observations of the archetypal supernova (SN) impostor SN 1997bs.  We show that SN 1997bs remains much fainter than its progenitor, posing a challenge for
the canonical picture of late-time obscuration by dust forming in a shell ejected during the transient.  The possibility that the star survived cloaked behind a dusty, steady wind is also disfavored.  The simplest explanation is that SN 1997bs was a subluminous Type IIn SN, although it is currently impossible to rule out the 
possibility that the star survived either behind an obscuring dusty shell $\gtrsim 1 M_{\odot}$ or with a significantly decreased intrinsic luminosity.  Continued monitoring of the system is the most promising means of ultimately resolving this question.
\end{abstract}

\begin{keywords}
stars: evolution -- supergiants -- supernovae: general -- supernovae: individual: SN 1997bs
\end{keywords}

\section{Introduction}
Supernova (SN) impostors \citep{vanDyk00} are a class of stellar transients characterized by Type IIn spectra (narrow hydrogen emission lines) with lower peak luminosities ($M_{V}\simeq-13$) than typical core collapse SNe (ccSNe; $M_{V}\simeq-17$).  The class is heterogeneous, including luminous variable stars \citep[e.g. SN 2002kg;][]{Weis05} and the SN 2008S class of transients \citep{Prieto08,Thompson09,Kochanek11b}, but a subset of the events are generally connected to some type of eruptive transient associated with the phenomenology of luminous blue variable stars \citep[LBVs;][]{Humphreys94}.  Still, the true nature of SN impostors is debated, including whether they are non-terminal eruptions or actual SNe \citep[see][]{Smith11,Kochanek12}.

The rate of SN impostors attributed to LBV eruptions is likely $\sim20\%$-$60\%$ of the ccSN rate \citep{Thompson09}.  Thus, the true nature of these transients could have important consequences for the rate of SNe.  For example, the SN rate appears to be less than the massive-star formation rate \citep{Horiuchi11}.  There are several possible solutions to this mismatch.  Non-local SN surveys could be significantly incomplete \citep[e.g.,][]{Botticella12} or there could be a significant rate of failed SNe \citep{Kochanek08,Gerke15}.  The latter solution could also help to explain the lack of higher mass SN progenitors \citep[$\gtrsim15~M_{\odot}$;][]{Kochanek08,Smartt09} and the black hole mass function \citep{Kochanek14b,Kochanek15}.  The final option is that some of these lower luminosity transients are in fact SNe \citep{Horiuchi11}.

On the other hand, if SN impostors are non-terminal events, they may be the dominant mode of mass-loss in massive stars \citep{Smith06}.  
This would be important because eruptive mass-loss is unaccounted for in stellar evolution models and little is known about the mechanism, duty cycle, mass dependence, or total mass-loss in such events.  Recently \cite{Khan13,Khan15} surveyed nearby galaxies and found that luminous stars encased in ejected dusty shells analogous to $\eta$ Carinae are not common enough for eruptions to be the dominant mass-loss channel, representing $\sim10\%$ of mass-loss rather than $\sim50\%$. 

Inspired by the phenomenology of $\eta$ Carinae, the most-widely accepted picture of SN impostors is that these events are the result of non-terminal eruptions that eject a significant amount of mass that then may obscure the surviving star \citep[see e.g.,][]{Humphreys99}.  The phenomenology expected for a dusty shell is fairly simple -- the optical depth peaks as the shell passes through the dust formation radius and then declines as $\sim t^{-2}$ with time \citep{Kochanek11}.  \cite{Kochanek12} examined several SN impostors with archival \emph{Hubble Space Telescope} (\emph{HST}), \emph{Spitzer Space Telescope} (\emph{SST}), and Large Binocular Telescope (LBT) data and found that none of the transients (for which there are sufficient data to reach a conclusion) are consistent with the shell ejection scenario.
If they are simply outbursts, the data are more consistent with a scenario where the transient is a sign post that the star is transitioning from a low to a high mass-loss rate, resulting in a longer lived dusty wind.  Still, nothing in the available data ruled out the possibility that some of these systems are not impostors at all, but instead are true SNe.

The basic problem is that SN impostors have been little studied after the brief optical transient even though it is really the late-time phenomenology that is the key to understanding these events.  For example, it would be useful to simply establish the continued presence of a star.
In this paper we utilize new late-time observations to investigate the nature of the archetypal impostor SN 1997bs.

SN 1997bs, discovered on 1997 April 15, was the first SN reported by the Lick Observatory Supernova Search \citep[LOSS;][]{Treffers97}.  It was found in NGC 3627 and was classified as a Type IIn SN (the spectrum was dominated by Balmer emission lines with full width at half-maximum $\simeq 1000~\mathrm{km}\>\mathrm{s}^{-1}$).  
\cite{vanDyk99} identified a candidate progenitor with $m_{F606W}\simeq22.86\pm0.16$ mag, which corresponds to a luminosity of $L_{*}=10^{4.8}$--$10^{5.4}~L_{\odot}$ for the temperature range $T_{*}=7500$--20,000 K.
The event peaked at $V \simeq 17$ and had faded to ($F555W$) $V \simeq 23.4$ mag by 1998 January 10.  Over this period the transient became redder, evolving from $V-I=0.7$ to $V-I=3$ mag \citep{vanDyk00}.  \cite{vanDyk00} suggested that the flattening of the late-time light curve at $\sim 0.5$ mag fainter than the progenitor was an indication that the star survived the explosion.

\begin{figure}
  \ifpdflatex
    \includegraphics[width=8.6cm, angle=0]{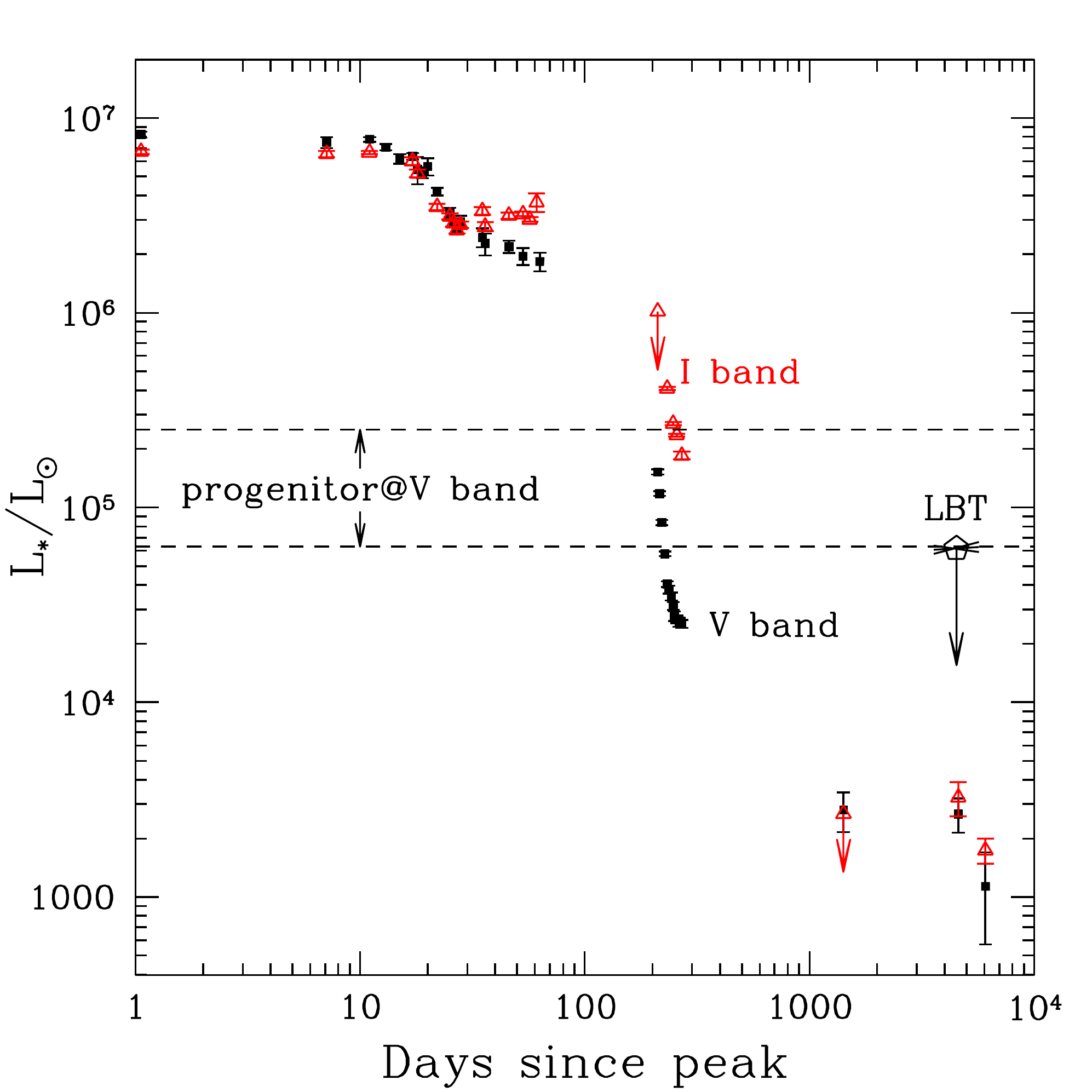}
  \else
    \includegraphics[width=8.6cm, angle=0]{../../lc.ps}
  \fi
  \caption{Optical light curve of SN 1997bs. The black filled squares (red open triangles) show the evolution of the \emph{V}-band (\emph{I}-band) luminosity
 converted from the \emph{HST} magnitudes reported by \citet{vanDyk00} and \citet{Li02} as well as our measurements of archival \emph{HST} images from 2004 and 2009.  The last set of \emph{V} and \emph{I}-band luminosities represent the upper limits we find for the new \emph{HST} data we present in this paper.  The short lines extending from the LBT point (the open pentagon) represent the \emph{V}-band variability and 3$\sigma$ limits found by image subtraction of our LBT monitoring data.  The horizontal dashed lines give the range of progenitor luminosities for $T_{*}=7500-20,000$ K given the candidate progenitor detection of $m_{F606W} \simeq 22.86\pm0.16$ mag reported by \citet{vanDyk99}. \label{fig:lightcurve}}
\end{figure}

In fact, it continued to fade, since \cite{Li02} found that the SN was marginally detected in an \emph{HST}/Wide Field Planetary Camera 2 (WFPC2) image taken in 2001, with $F555W$ = $25.8\pm0.3$ mag (March 4) and that it was not detected at $F814W$ to a limiting magnitude of about 25.0 (February 24 and May 28).  \cite{Li02} posited that while the formation of dust in the ejecta could explain the continuous decline in the optical flux, this would be inconsistent with the color evolution.  The SN would become progressively redder if dust was forming in the ejecta, but instead the source appeared to become bluer between early 1998 and early 2001.  We present the full optical light curve in Fig. \ref{fig:lightcurve}.

\cite{vanDyk12} reported mid-infrared emission near the position of SN 1997bs in \emph{SST} Infrared Array Camera (IRAC) images from 2004 May, with 32 $\mu$Jy at $3.6~\mu\mathrm{m}$ and 40 $\mu$Jy at $4.5~\mu\mathrm{m}$, but no detections at $5.8~\mu\mathrm{m}$ and $8.0~\mu\mathrm{m}$.  \cite{vanDyk12} stated that this could be fitted by a $T \simeq 970$ K blackbody with a radius $R \simeq 1.4 \times 10^{15}$ cm and luminosity 
$L \simeq 3.1 \times 10^{5}~L_{\odot}$.  This would correspond to an expansion speed of only $\sim60~\mathrm{km\>s^{-1}}$, far slower than the $\sim765~\mathrm{km\>s^{-1}}$ line width observed during the transient \citep{Smith11}.  Additionally, \cite{vanDyk12} found that the star is recovered at $m_{F555W} = 26.08$ and $m_{F814W} = 25.08$ mag in \emph{HST}/Advanced Camera for Surveys (ACS) data obtained in 2009, although they do not report uncertainties.

\cite{Kochanek12} considered the same \emph{SST} observations but had a different interpretation.  They found that most of the flux of the point source possibly present in the $4.5~\mu\mathrm{m}$ image vanished in wavelength-differenced images\footnote{Using difference imaging methods to subtract the 3.6 and $4.5~\mu\mathrm{m}$ images removes all sources with the `Rayleigh--Jeans' mid-IR spectral energy distributions of normal stars to leave only those with significant dust emission.  Thus, the lack of a counterpart in the wavelength-differenced images indicates an absence of dust emission.}, suggesting that the source of the flux was not dusty.
Furthermore, they also found that the evolution of the spectral energy distribution (SED) is inconsistent with the ejection of a single shell at the time of the transient.  
Keeping the star obscured at the time of the \emph{HST} observations in 2001 would require a continued mass-loss rate of $\dot{M} \sim 10^{-3} M_{\odot}\>\mathrm{yr}^{-1}$.  However, matching the candidate surviving star to the progenitor requires a temperature ($T_{*}>10^{4}$ K), too high to allow the formation of dust \citep[see][]{Kochanek11b,Kochanek14}.  \cite{Kochanek12} concludes that if the star found by \cite{Li02} is the survivor of SN 1997bs, the most likely scenario is that the eruption ended before 2001 and the star would have to be unobscured in 2011.

\begin{figure}
  \ifpdflatex
    \includegraphics[width=8.6cm, angle=0]{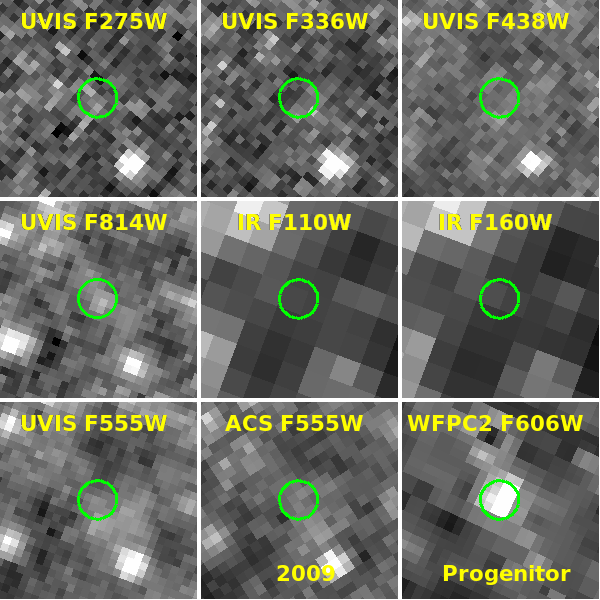}
  \else
    \includegraphics[width=8.6cm, angle=0]{../../data/N3627/wfc3_march/postage_stamps_labels3x3.eps}
  \fi
  \caption{The region surrounding SN 1997bs in the new \emph{HST} WFC3 images, the ACS $F555W$ image from 2009, and the pre-SN WFPC2 $F555W$ image.   The radius of the green circles (0$\farcs$1) is 25 times our positional uncertainty and each image is $1\farcs0$ across. \label{fig:postage_stamps}}
\end{figure}

In this work, we revisit these questions with new \emph{HST}, \emph{SST}, and LBT observations.  In \S\ref{sec:dataandmodels} we present the data, explain our identification of the source and its photometry, and present the methods we use to model and constrain the existence of a surviving star.  As can be seen from Fig. \ref{fig:postage_stamps}, no star is obviously visible, and there is certainly no source with a flux comparable to the progenitor.  
In \S\ref{sec:models} we present the results of modeling the SED of the source as obscured by an expanding shell or a steady-state wind and show that they are inconsistent with a surviving star as luminous as the progenitor.  Finally in \S\ref{sec:conclusions} we summarize our results and consider alternative explanations for the fate of SN 1997bs.
When converting observables into physical quantities we adopt the Cepheid distance of 9.4 Mpc to NGC 3627 from \cite{Freedman01} and a Galactic extinction of $E(B-V) = 0.04$ from the \cite{Schlafly11} recalibration of \cite{Schlegel98}.  We note that this results in a significantly smaller progenitor luminosity than that adopted by \cite{vanDyk00}, who used an older Cepheid distance of 11.1 Mpc from \cite{Saha99} and also included an estimate of galactic reddening from NGC 3627 of $E(B-V) = 0.21$ mag.  Though we do not include this galactic reddening, such reddening, to first order, will be taken into account by our circumstellar dust models \citep[but see][]{Kochanek12}.  Additionally, most of our results are dependent primarily on the luminosity of the progenitor relative to a possible survivor, which is independent of foreground extinction and the adopted distance.

\section{Data and Models}
\label{sec:dataandmodels}

\subsection{Data}
\label{sec:data}
We utilize both new and archival \emph{HST} data.  
For our program, we obtained new Wide Field Camera 3 (WFC3) UVIS $F555W$ and $F814W$ and IR $F110W$ and $F160W$ images (GO-13477) taken in 2013 November.  We also use public \emph{HST} WFC3 UVIS $F275W$, $F336W$, $F438W$, $F555W$, and $F814W$ images taken for the Legacy ExtraGalactic UV Survey (LEGUS; PI D. Calzetti, GO-13364) in 2014 February and the multi-epoch archival WFPC2 $F555W$ images of NGC 3627 taken between late 1997 and early 1998 (PI A. Sandage, GO-6549) that contain the transient event to calibrate our astrometry and measure the position of SN 1997bs.  Additionally, we analyze archival ACS/Wide Field Channel (WFC) $F555W$ and $F814W$ images taken in 2009 December (PI S. van Dyk, GO-11575) and $F435W$ images taken in 2004 December (PI R. Chandar, GO-10402).

We also make use of new and archival \emph{SST} data.  We co-added archival images of NGC 3627 (from program ID 159) taken in 2004 May and also co-added archival images taken in 2014 Feb-March-Aug (ID 10136) together with data from our program (ID 10001) taken in 2013 July using the {\sc mopex} reduction package\footnote{http://irsa.ipac.caltech.edu/data/SPITZER/docs/dataanal\-ysistools/tools/mopex/}.  We also performed image subtraction of these data using {\sc isis} \citep{Alard98,Alard00} to check for source variability.

We have been monitoring NGC 3627 with the LBT as part of a program searching for failed SNe \citep{Kochanek08,Gerke15}.  We use {\sc isis} to check for optical source variability in the dozen epochs we have collected between 2008 and 2014.

\subsection{Candidate Identification}
\label{sec:candidate}
Proper alignment of all the data and a precise measurement of the coordinates of SN 1997bs are critical for correct candidate identification and photometry.  We first aligned and stacked the archival drizzled \emph{HST} WFPC2 $F555W$ images using {\sc SExtractor} \citep{Bertin96}, {\sc scamp} \citep{Bertin06}, and {\sc swarp} \citep{Bertin02}.  We chose to register the astrometry of all our data to the drizzled \emph{HST} WFC3 $F814W$ image from 2013 November.  We found the position of SN 1997bs in our reference WFC3 $F814W$ frame by aligning the stacked archival WFPC2 $F555W$ image with the GEOMAP task in {\sc iraf}\footnote{{\sc iraf} is distributed by the National Optical Astronomy Observatory, which is operated by the Association of Universities for Research in Astronomy (AURA) under cooperative agreement with the National Science Foundation.} using a matched coordinate list of several dozen sources within 20$''$ of SN 1997bs.  We estimate an uncertainty of 0$\farcs$002 in the astrometry by measuring the rms in the SN position using different subsamples of the matched coordinate list with different orders for the astrometric fits.  We measure a centroid uncertainty of 0$\farcs$004 based on the aperture photometry of SN 1997bs in the stacked archival frame.  Adding these uncertainties in quadrature gives a total positional uncertainty of 0$\farcs$004 (0.1 pixels) in our reference image.

We generate photometric catalogs for the \emph{HST} WFC3 data using the software package 
\ifapj
  {\sc dolphot 2.0} \citep{Dolphin00}\footnote{\url{http://americano.dolphinsim.com/dolphot/}}.
\else
  {\sc dolphot 2.0} \citep{Dolphin00}\footnote{http://americano.dolphinsim.com/dolphot/}.
\fi
  We largely used the same parameter settings as in \cite{Dalcanton12}, with the exceptions listed in Table \ref{tab:dolphot}.
  We again used the drizzled \emph{HST} WFC3 $F814W$ image from 2013 November as the reference.  Although {\sc dolphot} has a built-in routine for aligning images, it was necessary to calculate initial pixel offsets using the {\sc tweakreg} task in the {\sc drizzlepac} package for the images taken on a different epoch (2014 February) than the reference.  The rms alignment residuals for the different images ranged from 0$\farcs$003 for the UVIS $F555W$ and $F814W$ images to 0$\farcs$03 for the UV images, and 0$\farcs$16 for the IR images.
We also performed aperture photometry with the PHOT task in {\sc iraf}.  The aperture corrections were calculated using the \emph{HST} point spread function (PSF) models from 
\ifapj
  {\sc tiny tim} \citep{Krist95,Krist11}\footnote{\url{http://tinytim.stsci.edu/cgi-bin/tinytimweb.cgi}}.
\else
  {\sc tiny tim} \citep{Krist95,Krist11}\footnote{http://tinytim.stsci.edu/cgi-bin/tinytimweb.cgi}.
\fi
\input{table1.tex}

The closest source found by {\sc dolphot} at \emph{I}-band is 0$\farcs$04 from the expected position of SN 1997bs with a centroid uncertainty estimated from the reported S/N of the detection of 0$\farcs$004.  We also estimate the uncertainty by bootstrap resampling the set of six $F814W$ images that are co-added to form the reference image.  The rms in the location of the closest source to the position of SN 1997bs using these different reference images is 0$\farcs$11.  The rms of the source location found using {\sc dolphot} with slightly different sets of parameters is $\sim 0\farcs03$.  Taken together, these uncertainties mean that the source position is consistent with the position of SN 1997bs.  The photometry for the ACS/WFC images was found by again running {\sc dolphot} with the same reference image.  We report the {\sc dolphot} photometry of the \emph{HST} data in Table \ref{tab:photometry}, separately listing the statistical uncertainties from {\sc dolphot} and systematic uncertainties estimated from the standard deviations of the fluxes measured using a range of {\sc dolphot} parameter settings.

We also looked for an IR source at the position of SN 1997bs.
First we registered the \emph{SST} images to the \emph{HST} data using the GEOMAP task in {\sc iraf} with coordinates of matched point sources.  The point sources were identified by subtracting \emph{SST} images convolved with a larger kernel from those convolved with a smaller kernel.  The rms in the astrometric solution is $\sim0\farcs12$.  There is no clear point source at the position of SN 1997bs in the \emph{SST} data but we give upper limits based on aperture photometry in Table \ref{tab:photometry}.  Unfortunately SN 1997bs is coincident with diffuse IR emission from a spiral arm, making photometric measurements challenging.  In order to minimize contamination from diffuse emission and other sources we use a $1\farcs2$ radius aperture with a $1\farcs2 - 3\farcs6$ radius sky annulus together with empirically-determined aperture corrections.  While such a small sky annulus includes significant point source flux, the background is non-uniform on larger scales.
We also measure the variability of SN 1997bs in the \emph{SST} images taken between 2004 and 2014 using image subtraction.  While the $3.6~\mu\mathrm{m}$ flux stays constant, the $4.5~\mu\mathrm{m}$ flux declines by 24 $\mu$Jy.

\newcommand{\photdata}{
    HST WFC3/UVIS $F275W$ & $>24.7$ & 2014-02-08 \\
    HST WFC3/UVIS $F336W$ & $26.72 \pm 0.69 \pm 0.63$ & 2014-02-08 \\
    HST WFC3/UVIS $F438W$ & $27.13 \pm 0.35 \pm 0.54$ & 2014-02-08 \\
    HST WFC3/UVIS $F555W$ & $26.90 \pm 0.15 \pm 0.52$ & 2013-11-28, 2014-02-08 \\
    HST WFC3/UVIS $F814W$ & $25.49 \pm 0.12 \pm 0.11$ & 2013-11-28, 2014-02-08 \\
    HST WFC3/IR $F110W$ & $23.83 \pm 0.04 \pm 0.35$ & 2013-11-28 \\
    HST WFC3/IR $F160W$ & $22.64 \pm 0.04 \pm 0.26$ & 2013-11-29 \\
    SST IRAC $3.6~\mu\mathrm{m}$ & $>18.2$ ($<15~\mu$Jy) & July 2013, Feb-Mar-Aug 2014 \\        
    SST IRAC $4.5~\mu\mathrm{m}$ & $>18.5$ ($<7~\mu$Jy) & July 2013, Feb-Mar-Aug 2014 \\ 
    SST IRAC $5.8~\mu\mathrm{m}$ & $>15.1$ ($<109~\mu$Jy) & May 2004 \\                  
    SST IRAC $8.0~\mu\mathrm{m}$ & $>12.5$ ($<645~\mu$Jy) & May 2004 \\                  
    SST IRAC $3.6~\mu\mathrm{m}$ & $>18.2$ ($<15~\mu$Jy) & May 2004 \\           
    SST IRAC $4.5~\mu\mathrm{m}$ & $>16.9$ ($<31~\mu$Jy) & May 2004 \\           
    HST ACS/WFC $F435W$ & $26.78 \pm 0.17 \pm 0.04$ & 2004-12-31 \\
    HST ACS/WFC $F555W$ & $25.97 \pm 0.17 \pm 0.03$ & 2009-12-14 \\
    HST ACS/WFC $F814W$ & $24.81 \pm 0.10 \pm 0.00$ & 2009-12-14 \\
    HST WFPC2 $F555W$ & $25.8 \pm 0.3$ 
      \ifapj
        \tablenotemark{a}
      \else 
        $^{a}$
      \fi
      & 2001-03-04 \\
    HST WFPC2 $F814W$ & $>25.0$ 
      \ifapj
        \tablenotemark{a}
     \else
       $^{a}$
     \fi
     & 2001-02-24, 2001-05-28 \\

}
\ifapj
  \begin{deluxetable*}{ccc}
  \tablecaption{Photometry}
  \tablehead{
    \colhead{Filter} &
    \colhead{Magnitude ($\pm$ stat $\pm$ sys)} &
    \colhead{Epoch} }
  \startdata
\else
  \begin{table*}
  \begin{minipage}{10cm}
  \caption{Photometry}
  \begin{tabular}{ccc}
  \hline
  \hline
  {Filter} & {Magnitude ($\pm$ stat $\pm$ sys)} & {Epoch} \\
  \hline
\fi
\photdata
\ifapj
  \enddata
  \tablenotetext{a}{Photometry taken from \cite{Li02}}
  \tablecomments{Upper limits are from aperture photometry.}
  \label{tab:photometry}
  \end{deluxetable*}
\else
  \hline
  \hline
  \end{tabular}
  $^{a}$Photometry taken from \cite{Li02} \\
  Magnitudes listed with uncertainties are from {\sc DOLPHOT} PSF photometry.  Magnitudes with only upper limits are from aperture photometry.
  \label{tab:photometry}
  \end{minipage}
  \end{table*}
\fi

We use the LBT data to place limits on optical variability at the coordinates of SN 1997bs.  No source is detected at the position of SN 1997bs and with a dozen epochs spanning 2008-2014 we find slopes consistent with zero in all filters
($150 \pm 420$, $-260 \pm 450$, $320 \pm 390$, and $310 \pm 660$
 L$_{\odot}\>\mathrm{yr}^{-1}$ and with rms residuals of 
1400, 3200, 1600, and 2900 L$_{\odot}$ in \emph{U}, \emph{B}, \emph{V}, and \emph{R} respectively, where the uncertainties include both statistical uncertainties and systematic uncertainties estimated from the rms of light curves of a sigma-clipped grid of points within 45 arc-sec of the coordinates of SN 1997bs).  The new \emph{HST} observations reveal that the $F555W$ luminosity at the position of 97bs decreased by $\sim1.1\pm0.6$ mag between 2001 and 2013 and the $F814W$ luminosity decreased by $\sim0.7\pm0.2$ mag between 2009 and 2013.  In principle, the \emph{HST} data can place more stringent limits on variability of the source due to the longer baseline and better resolution, but in practice the LBT variability limits may be more reliable and less susceptible to systematic errors by virtue of coming from a single instrument and using image subtraction. 

\subsection{Confusion}
\label{sec:confusion}
Since we are considering a faint source in a crowded field, we must evaluate the likelihood that our detected \emph{HST} source could be an incidental detection of an unrelated source.  We find that the surface density of all {\sc dolphot} sources within 0$\farcs$8
 of the position of SN 1997bs is 31/arc-sec$^{2}$, which corresponds to a $15\%$ chance of an unrelated source being detected within 0$\farcs$04 (the distance of the closest source from the SN location) by chance.  For sources as bright (in $F814W$) as the detection, the surface density is 7/sq-arcsec, which reduces the chance of confusion to $4\%$.  However, aperture photometry seems to indicate that confusion is more substantial, with $\sim 30\%$ of apertures laid out over a 0$\farcs$8 grid surrounding the source yielding fluxes as bright as our detection.  The large difference between the two methods might be due to {\sc dolphot} limiting detections to only nearly point-like objects or from source deblending.  

Another possibility is that the detection is a surviving companion to the progenitor of SN 1997bs.  Massive stars have a large multiplicity fraction
($>82\%$) \citep{Chini12}.
\cite{Kochanek09} estimates the magnitude distribution of surviving companions of ccSNe for a set of progenitor masses assuming a uniform distribution of mass ratios.  Unfortunately, the mass of the progenitor of SN 1997bs is not well constrained, but out of the four progenitor models calculated, the progenitor's $m_{F555W}=22.86$ is most consistent with the $20 M_{\odot}$ model.  For this progenitor mass and assuming a binary fraction of $80\%$ there is very roughly a 5-$20\%$ chance that there would be a surviving companion to SN 1997bs brighter than our detection.  However, as discussed in \cite{Kochanek09}, most secondaries of exploding stars are fainter, blue main sequence stars, but our detection is relatively red ($V-I\sim1.4$) 
and would correspond to $T_{*}\sim4400$ K
 if the SED is not strongly influenced by dust.  This reduces the likelihood that the detection is a surviving companion.

For the \emph{SST} data the variability in the $4.5~\mu\mathrm{m}$ flux strongly suggests that the $4.5~\mu\mathrm{m}$ flux observed in 2004 is associated with SN 1997bs.  The IR flux remaining in 2013/14 may be only due to diffuse emission coincident with the original source.

\subsection{Basic Scalings for Dust}

A common picture of SN impostors is that a surviving star is obscured by a dusty shell ejected at the time of the transient.  For such a model, geometry dictates that at late times the evolution of the optical depth of a uniform shell expanding at a constant velocity is $\tau(t) \propto t^{-2}$, where $t$ is the time elapsed since the ejection of the shell.  The observed visible light is dominated by scattered photons rather than direct emission, so inhomogeneities in the shells have less effect than naively expected \citep[see][]{Kochanek12}.  In particular, inhomogeneities can only grow with time and accelerate the optical depth evolution, which would further strengthen our arguments.  The evolution of the optical depth in turn must result in a change in the observed luminosity (in a given filter).  Consequently, the limits on the observed variability of a source surrounded by an expanding shell can be used to constrain the current effective absorption optical depth, $\tau_{\mathrm{eff}} = \left[ \tau_{\mathrm{abs}} (\tau_{\mathrm{abs}} + \tau_{\mathrm{sca}}) \right]^{1/2}$, where $\tau_{\mathrm{abs}}$ and $\tau_{\mathrm{sca}}$ are the absorption and scattering optical depths, by
\begin{equation}
\tau_{V,\mathrm{eff}}<\frac{1}{2}\frac{t}{L_{V,\mathrm{obs}}}\left(\frac{dL_{V,\mathrm{obs}}}{dt}\right) ,
\label{eqn:taulimit}
\end{equation}
where $t$ is again the elapsed time, $L_{V,\mathrm{obs}}$ is the current observed luminosity, and $dL_{V,\mathrm{obs}}/dt$ is the limit on the rate of change in the observed \emph{V}-band luminosity.
Similarly, variability limits constrain the maximum luminosity, $L_{*,V}$, of a star surviving within an expanding shell to
\begin{equation}
L_{*,V}< \frac{1}{2}\frac{t}{\tau_{V,\mathrm{eff}}}\left(\frac{dL_{V,\mathrm{obs}}}{dt}\right) \mathrm{e}^{\tau_{V,\mathrm{eff}}} .
\label{eqn:lumlimit}
\end{equation}
Late-time data, in addition to constraining the optical depth of an expanding shell and the luminosity of a surviving star from the observed variability, can also be used to estimate the mass of the shell.
The mass of the shell, $M_{\mathrm{ej}}$, is related to the total optical depth, $\tau_{V,\mathrm{tot}}$, by
\begin{equation}
M_{\mathrm{ej}} = \frac{ 4 \pi v_{\mathrm{e}}^{2} t^{2} \tau_{V,\mathrm{tot}}(t)}{\kappa_{V}} ,
\label{eqn:ejectedmass}
\end{equation}
where $v_{\mathrm{e}}$ is the radial velocity of the shell and $\kappa_{V}$ is the opacity at \emph{V}-band.  The total and effective optical depths are related by the albedo $w$ with $\tau_{\mathrm{eff}} = (1-w)^{1/2} \tau_{\mathrm{tot}}$.

Another case to consider is that the progenitor star is obscured by a steady-state dusty wind.  We can estimate the mass-loss rate needed to obscure the progenitor star with such a wind.  If we assume that all of the dust forms at the dust formation radius, $R_{\mathrm{f}}$, then the rate of mass-loss is:
\begin{equation}
\dot{M} = \frac{4\pi v_{\mathrm{w}} R_{\mathrm{f}} \tau_{V,\mathrm{tot}}}{\kappa_{V}}
\label{eqn:mdot}
\end{equation}
where $R_{\mathrm{f}} \sim L^{1/2} / T_{\mathrm{f}}^{2}$ and $T_{\mathrm{f}} \sim 1500$ K is the dust formation temperature.
More precisely the dust begins to form at $R_{\mathrm{f}}$, where the temperature is low enough for grain condensation at a rate proportional to density.  As discussed in \cite{Kochanek11}, $R_{\mathrm{f}}$ has a complex temperature dependence, but would only range from $2.3\times 10^{14}$ cm for a $10^{4.7} L_{\odot}$, 7500 K star to $2.8\times 10^{15}$ cm for a $10^{5.4} L_{\odot}$, 20,000 K star.  The dust opacity depends on the distribution of grain sizes and the dust-to-gas ratio, but will generally be within a factor of 2 of 100 cm$^{2}$/g (of gas) at V-band.
If the dust is aspherically distributed, the ejected mass inferred by Eqn. \ref{eqn:ejectedmass} for the shell case and the rate of mass-loss for the wind case from Eqn. \ref{eqn:mdot} could be over- or underestimates depending on the specific geometry and orientation of the system.  However, the temporal scalings for the optical depth encoded by Eqn. \ref{eqn:taulimit} and \ref{eqn:lumlimit} should remain reasonable approximations even for complex geometries \citep{Kochanek12}.

We will utilize these relations in \S\ref{sec:shell} and \S\ref{sec:wind} to help determine whether a surviving star to SN 1997bs could be obscured by an expanding shell or steady-state wind.

\subsection{DUSTY}

We model the SED of the source using {\sc dusty} \citep{Ivezic97,Ivezic99,Elitzur01}, a code for solving radiative transfer through a spherically symmetric dusty medium.  We use stellar atmospheric models from \cite{Castelli04} for stars of various temperatures and solar composition.
We find best-fitting models using a Markov Chain Monte Carlo (MCMC) wrapper around {\sc dusty} for both silicate and for graphitic dust from \cite{Draine84} using a standard MRN grain size distribution \cite[$dn/da \propto a^{-3.5}$ with $0.005~\mu\mathrm{m} < a < 0.25~\mu\mathrm{m}$;][]{Mathis77}.  In Appendix \ref{app:one} we show that our conclusions are not dependent on this choice for the grain size distribution.  Since silicate dust is the type expected to form around more massive stars (like the progenitor of SN 1997bs was believed to be) and the conclusions of the paper are robust to the chosen dust type, we will only present the results for silicate dust.
For the MRN grain size distribution of silicate dust $w_{V} \simeq 0.86$, making $\tau_{V,\mathrm{eff}} \simeq 0.37 \tau_{V,\mathrm{tot}}$.
The models assume that a shell with a thickness ($R_{\mathrm{out}}/R_{\mathrm{in}}$) of 2.0 expanding with a velocity of $v_{\mathrm{e}} = 765~\mathrm{km\>s^{-1}}$ \citep[$\pm$ a factor of 2;][]{Smith11} was ejected during the 1997 event.  We actively expand $R_{\mathrm{in}}$ at this rate since the effects of dust are dominated by $R_{\mathrm{in}}$ because optical depth drops as $R^{-1}$.  We also produce models for a dusty wind by setting $R_{\mathrm{in}}$ equal to the radius corresponding to a typical dust formation temperature of 1500 K \citep[see, e.g.,][]{Kochanek14}.
The $\chi^{2}$ used in the MCMC is calculated with the logarithmic (linear) differences between modeled and observed fluxes (or limits) in each filter treated as a detection (or limit) and, for the expanding shell models, between modeled and observed $v_{\mathrm{e}}$ (to place the dust at the $R_{\mathrm{in}}$ corresponding to the given dust temperature).  
When considering all photometric constraints as only upper limits, we calculate the $\Delta\chi^{2}$ compared to having no star as functions of $L_{*}$ and $T_{*}$.  When folding in the light curves we add $\chi^2$ contributions for each observation, $n$, in each filter, $f$, found by 
\begin{equation}
\chi^2_{f,n} = \left( \frac{L_{\mathrm{obs},f,n} - L_{\mathrm{mod},f,n}}{\sigma_{L_{\mathrm{obs},f,n}}} \right)^2 ,
\label{eqn:lumchi2}
\end{equation}
where the model luminosity
\begin{equation}
L_{\mathrm{mod},f,n} = L_{*,f} \mathrm{e}^{-\tau_{\mathrm{eff},f,n}} 
\label{eqn:lummodel}
\end{equation}
is controlled by the evolution of the optical depth
\begin{equation}
\tau_{\mathrm{eff},f,n} = \tau_{\mathrm{eff},f} \left[ \left( \frac{T}{t_{n}} \right)^2 - 1 \right] .
\label{eqn:tauevol}
\end{equation}
In this model, $\tau_{\mathrm{eff},f}$ is the optical depth at the time of the latest observations $T=16.5$ yr after the transient and $t_{n}$ is the time elapsed since the transient for observation $n$.

\begin{figure}
  \begin{center}
  \ifpdflatex
    \includegraphics[width=8.6cm]{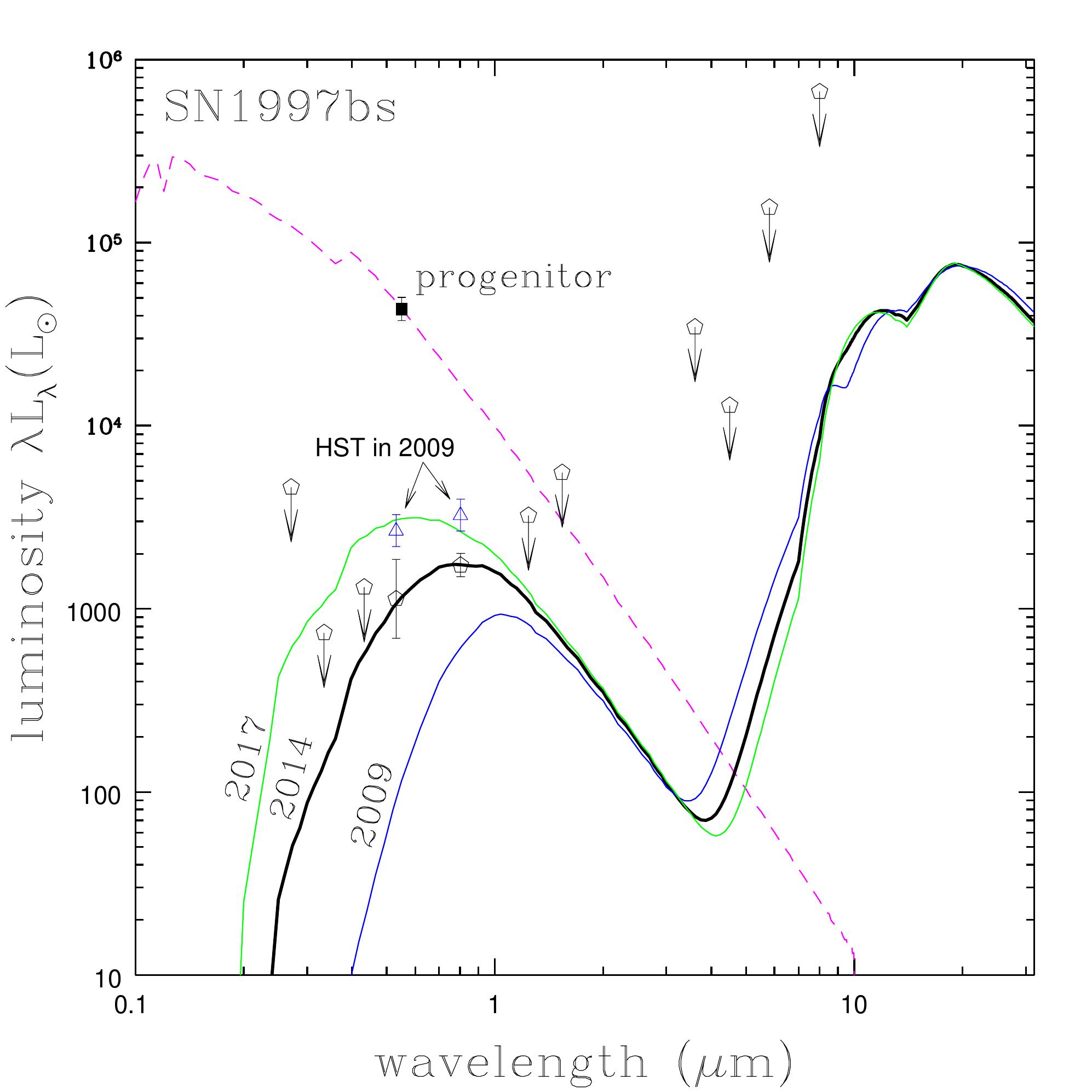}
  \else
    \includegraphics[width=8.6cm]{../../dusty/mylimits/2014_silicate_shell_f814f555_mcmc/sed_extrapolation.ps}
  \fi
  \end{center}
  \caption{The SED of SN 1997bs with the best-fitting model for a silicate shell when treating our $F555W$ and $F814W$ photometry as detections and the other bands as upper limits.  The latest \emph{HST} and \emph{SST} detections and limits for a surviving star are shown as the open pentagons with the current best-fitting observed model spectrum shown as the thick black line.  This best-fitting model has $T_{*}=$ 18,900 K, $\tau_{V,\mathrm{tot}}=5.34$, and $L_{*}=10^{4.90}~L_{\odot}$.  We also show this best-fitting model evolved forwards in time to 2017 in green and backwards in time to 2009 in blue with $\tau \propto t^{-2}$ but with the same $T_{*}$ and $L_{*}$.  The tension between the 2009 \emph{HST} photometry, shown as the open blue triangles, with the best-fitting shell model evolved back to 2009 illustrates how models in which SN 1997bs is obscured by an expanding dusty shell cannot be reconciled with the late-time optical evolution.  Although the present-day luminosity constraints allow for SED solutions with higher IR fluxes these solutions require higher optical depths, which, when evolved backwards in time, are even more incompatible with the 2009 \emph{HST} photometry.  For comparison, we show the \emph{HST} progenitor detection as the solid black square and an unobscured model spectrum with $T_{*}=$ 18,900 K and $L_{*}=10^{5.47}~L_{\odot}$ as the dashed magenta line. \label{fig:sed_shell}}
\end{figure}

\section{A Star Obscured by Dust?}
\label{sec:models}
We consider whether the progenitor of SN 1997bs could have survived obscured by either an expanding dusty shell or by a steady-state wind.  Our candidate late-time detection, like those in earlier studies, is marginal and confusion is a non-trivial issue.  Therefore, we will consider a range of interpretations for which observations constitute detections and which are only upper limits.  We consider the case of $F555W$ and $F814W$ as detections since detections in these filters have been previously reported \citep{Li02,vanDyk12}.  We also consider the case of only $F814W$ as a detection, as this is the filter in which a detection is most convincing visually.  We consider the case of $F814W$ and $4.5~\mu$m as detections because there was a decline in $4.5~\mu$m flux between 2004 and 2014 and this represents a scenario in which there is high obscuration.  Finally, we consider the case where all of the photometry is treated as upper limits, since the candidate detections are marginal and could be due to confusion.

\subsection{Obscuration by an Expanding Shell}
\label{sec:shell}
First, we model our photometry as though a surviving star obscured by an expanding shell is recovered with detections in $F555W$ and $F814W$ but only upper limits in the other filters.  We present, as an example, the best-fitting SED for this case in Fig. \ref{fig:sed_shell}.  This best-fitting SED is almost a factor of 4 fainter than the luminosity of a progenitor with the same $T_{*}$ and relies on a large amount of obscuration ($\tau_{V,\mathrm{tot}}=5.3$) to account for the low optical flux currently observed.

\begin{figure*}
  \begin{center}
  \ifpdflatex
    \includegraphics[width=0.78\textwidth]{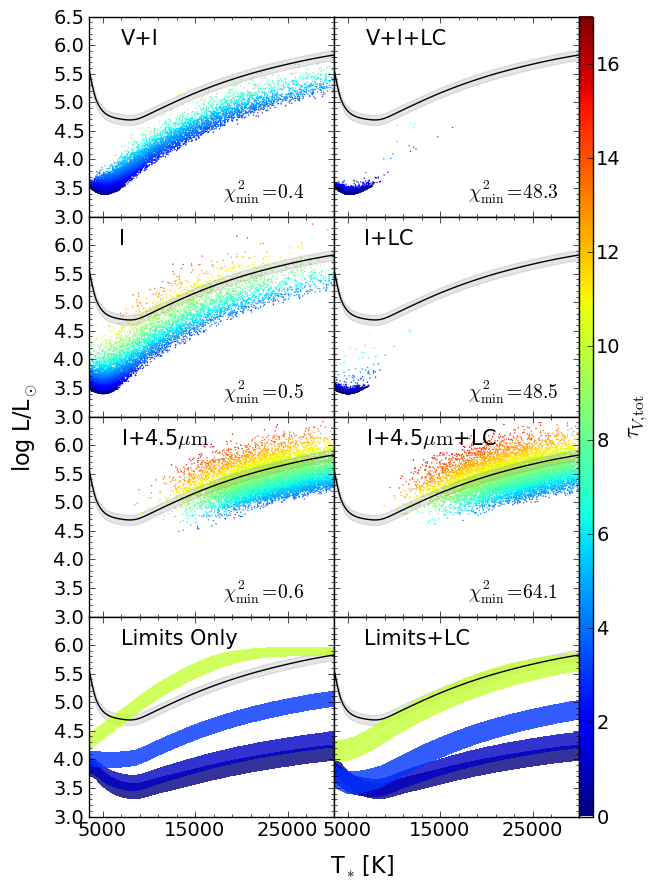}
  \else
    \includegraphics[width=0.78\textwidth]{../../dusty/mylimits/fig4.ps}
  \fi
  \end{center}
  \caption{MCMC results for the luminosity and temperature of a surviving star obscured by an expanding shell ejected at the time of the SN 1997bs transient when imposing different possible photometric constraints: treating the \emph{HST} WFC3/UVIS $F555W$ and $F814W$ photometry as detections and the other filters as only upper limits (top row), treating the \emph{HST} WFC3/UVIS $F814W$ photometry as a detection and the other filters as upper limits (2nd row from the top), treating the \emph{HST} WFC3/UVIS $F814W$ and \emph{SST} $4.5~\mu$m photometry as detections and the other filters as upper limits (2nd row from the bottom), and treating all photometry as upper limits (bottom row).  The panels on the right also fold in the constraints from the LBT and \emph{HST} light curves.  
The shaded bands in the panels on the bottom row show the luminosities within the 90-99.99\% confidence intervals ($6.25 < \Delta \chi^2 < 21.1$ for three parameters -- $L_{*}$, $T_{*}$, and $\tau$) relative to no surviving star) for $\tau_{V,\mathrm{tot}}=$ 0, 1, 3, and 10.
For comparison, the solid black line and gray band indicates the progenitor luminosity as constrained by the pre-explosion measurement of $m_{\mathrm{F606}}=22.86$ and its $1\sigma$ uncertainty (0.16 mag).  The lowest $\chi^2$ value for the 10,000 MCMC steps accepted is given for each panel.  The panels on the left show that the latest photometric constraints taken alone do not rule out the star surviving with its pre-eruption luminosity if obscured by a shell with $\tau_{V,\mathrm{tot}} \gtrsim 5$.  
However, the panels on the right that fold in the light curve evolution only allow a luminous surviving star in the `\emph{I}$+4.5\mu\mathrm{m}+$LC' and `Limits$+$LC' cases.  The former is heavily disfavored by the relatively high $\chi^2_{\mathrm{min}}$ while the latter would require that all late-time optical detections are of an unrelated source.
\label{fig:shell}}
\end{figure*}

The complete results of our MCMC modeling are shown in Fig. \ref{fig:shell}.  When considering only the latest photometric constraints for each filter (the cases displayed on the left side of Fig. \ref{fig:shell}), it is the possible that the star survived with unchanged luminosity but is cloaked beneath a significant amount of extinction.  In the cases where either the $F555W$ and $F814W$ photometry or only the $F814W$ photometry is treated as a detection, the star can be cool with relatively low luminosity and low obscuration or hot with higher luminosity and higher obscuration.  The $F814W$-only case allows a more luminous surviving star than the $F555W$ and $F814W$ case because the models no longer have to fit the shallow slope between the $F555W$ and $F814W$ magnitudes, enabling scenarios with higher optical depths to achieve good fits.  In the $F814W$ and $4.5~\mu$m case, a cooler star with lower luminosity and lower obscuration is not allowed by the mid-IR detection, which requires a significant luminosity to be reprocessed by dust.  If all of the photometry is taken as only upper limits, the MCMC modeling would be poorly constrained, so we instead show the maximum allowed luminosity of a surviving star for optical depths of $\tau_{V,\mathrm{tot}}=$ 0, 1, 3, and 10.  The limits-only case echoes the results of the other cases -- the latest photometric constraints, taken alone, allow for a luminous, but heavily obscured, surviving star. 

These limits are relatively robust to variations in the velocity of the expanding shell.  Doubling the assumed velocity of the expanding shell only significantly increases the maximum luminosity of a surviving star with $T_{*}\gtrsim25,000$ K and $\tau_{V,\mathrm{tot}}=10$ and then only by $\sim0.3$ dex.  Conversely, halving the expansion velocity decreases the maximum luminosity in the same parameter range by a similar amount.

That solutions consistent with the photometric data exist does not mean they are physical because they may imply unphysical ejecta masses or unobserved variability.  First, we consider the ejecta mass needed to obscure a surviving star with an unchanged intrinsic luminosity and temperature.  Fig. \ref{fig:shell} shows that this would require the current optical depth to be $\tau_{V,\mathrm{tot}} \sim 10$.  Using Eqn. \ref{eqn:ejectedmass}, we estimate
\begin{eqnarray}
M_{\mathrm{ej}} = 1.0 \left( \frac{v_{\mathrm{e}}}{765~\mathrm{km\>s^{-1}}} \right)^{2} \left( \frac{t}{16.5~\mathrm{yr}} \right)^{2} \nonumber \\ \times \left( \frac{\tau_{V,\mathrm{tot}}}{10} \right) \left( \frac{100~\mathrm{cm^{2}/g}}{\kappa_{V}} \right) M_{\odot} .
\label{eqn:scaledejectedmass}
\end{eqnarray}
Such an ejected mass is feasible, but, as we shall see, this scenario is inconsistent with the variability constraints dictated by geometric expansion if any of the late-time photometric detections are powered by the luminosity from a surviving star. 
The kinetic energy of the ejecta, $E_{\mathrm{kin}} = \frac{1}{2} M_{\mathrm{ej}} v_{\mathrm{e}}^{2}$,
in this scenario would be
\begin{eqnarray}
E_{\mathrm{kin}} = 6 \times 10^{48} \left( \frac{v_{\mathrm{e}}}{765~\mathrm{km\>s^{-1}}} \right)^{4} \left( \frac{t}{16.5~\mathrm{yr}} \right)^{2} \nonumber \\ \times \left( \frac{\tau_{V,\mathrm{tot}}}{10} \right) \left( \frac{100~\mathrm{cm}^{2}/\mathrm{g}}{\kappa_{V}} \right)~\mathrm{erg} .
\end{eqnarray}
We compare this to the radiated energy, $E_{\mathrm{rad}} = t_{1.5} \zeta L_{\mathrm{peak}}$, of the transient estimated by \cite{Smith11} of
\begin{equation}
E_{\mathrm{rad}} = 7 \times 10^{49} \left( \frac{t_{1.5}}{45~\mathrm{days}} \right) \left( \frac{L_{\mathrm{peak}}}{1.2 \times 10^{7}~L_{\odot}} \right) \zeta~\mathrm{erg} ,
\end{equation}
where $L_{\mathrm{peak}}$ is the peak luminosity of the outburst, $t_{1.5}$ is the time for the transient to fade by 1.5 mag from its peak, and $\zeta \sim 1$ is a dimensionless factor that depends on the shape of the light curve.  The high ratio of radiated to kinetic energy in this scenario would likely require a radiative, rather than explosive, mechanism.  We do caution that these estimates of $M_{\mathrm{ej}}$ and $E_{\mathrm{kin}}$ assume homogeneous, spherically distributed ejecta.

Fig. \ref{fig:sed_shell} also illustrates how the expanding shell model that best fits the latest data would necessarily have been much fainter in the optical in 2009, in gross disagreement with the \emph{HST} photometry from 2009.  A model in which $\tau_{V,\mathrm{tot}}=10$ would be in even more severe conflict with the photometric evolution.  More quantitatively, using Eqn. \ref{eqn:taulimit} with the 3$\sigma$ upper limit on the variability from the LBT data constrains the current maximum optical depth of an expanding shell to be
\begin{equation}
\tau_{V,\mathrm{eff}} < 11 \left(\frac{t}{16.5~\mathrm{yr}}\right) \left(\frac{1140~L_{\odot}}{L_{V,\mathrm{obs}}}\right) \left(\frac{dL_{V,\mathrm{obs}}/dt}{1500~L_{\odot}\>\mathrm{yr}^{-1}}\right) .
\end{equation}
Similarly, we can use the variability limits from LBT and Eqn. \ref{eqn:lumlimit} to constrain the maximum luminosity of a star surviving within an expanding shell as
\begin{equation}
L_{*,V}< 3.3 \times 10^{4} \left(\frac{t}{16.5~\mathrm{yr}}\right) \left(\frac{dL_{V,\mathrm{obs}}/dt}{1500~L_{\odot}\>\mathrm{yr}^{-1}}\right) \frac{\mathrm{e}^{\tau_{V,\mathrm{eff}}-1}}{\tau_{V,\mathrm{eff}}} L_{\odot} \label{eqn:lumlimitscaled} .
\end{equation}
These limits from the LBT are less constraining than the decline in flux seen in the \emph{HST} observations.  In Fig. \ref{fig:shell}, the panels on the right show the constraining effect of using the both the LBT and \emph{HST} light curves from 2008-14 with Eqn. \ref{eqn:lumchi2} and \ref{eqn:lummodel}.  The effects of the luminosity constraints should be relatively robust to departures from our assumptions of homogeneous, spherically-distributed dust.
The light curves restrict solutions to have low optical depths ($\tau_{V,\mathrm{tot}} \lesssim 1$).  Although the \emph{I}$+4.5~\mu\mathrm{m}+$LC case still appears to favor a luminous, obscured survivor, the $\chi^{2}_{\mathrm{min}}$ for this case is much higher than that for the other cases, indicating that it is strongly disfavored.
Moreover, the putative \emph{HST} source at the coordinates of 97bs seems to have faded between 1998 and 2013, which further disfavors any expanding shell scenario in which the optical flux at the position of 97bs is from a surviving star.  The remaining expanding shell scenario is that the star is still heavily obscured ($\tau_{V,\mathrm{tot}}\gtrsim10$) and the optical emission reported here and by \cite{Li02} and \cite{vanDyk12} is not from a surviving star.

Such a scenario is difficult to reconcile with the first year of the transient light curve.  The light curve presented by \cite{vanDyk00} shows that SN 1997bs rapidly evolved to the red, with the redward color evolution peaking at $V-I \sim 3$ by eight months after discovery, which given the constraints on the ejecta velocity and the progenitor \emph{V}-band detection is roughly the time at which the ejecta should have reached the dust formation radius.  As discussed in \cite{Kochanek12}, this peak in the color evolution corresponds to a peak in the optical depth at $\tau_{V}\simeq10$.  A shell with $\tau_{V}=10$ at eight months would have evolved to $\tau_{V}\sim 0.02$ by our latest \emph{HST} observations.  Such a low optical depth would not allow a luminous survivor.  It may be possible that slower-moving ejecta reached the dust formation radius and formed additional dust during gaps in the coverage of the light curve later than nine months post-discovery, but all subsequent observations have constrained SN 1997bs to be much less red than it was at eight months.

\subsection{Obscuration by a Wind}
\label{sec:wind}
\begin{figure}
  \ifpdflatex
    \includegraphics[width=8.6cm, angle=0]{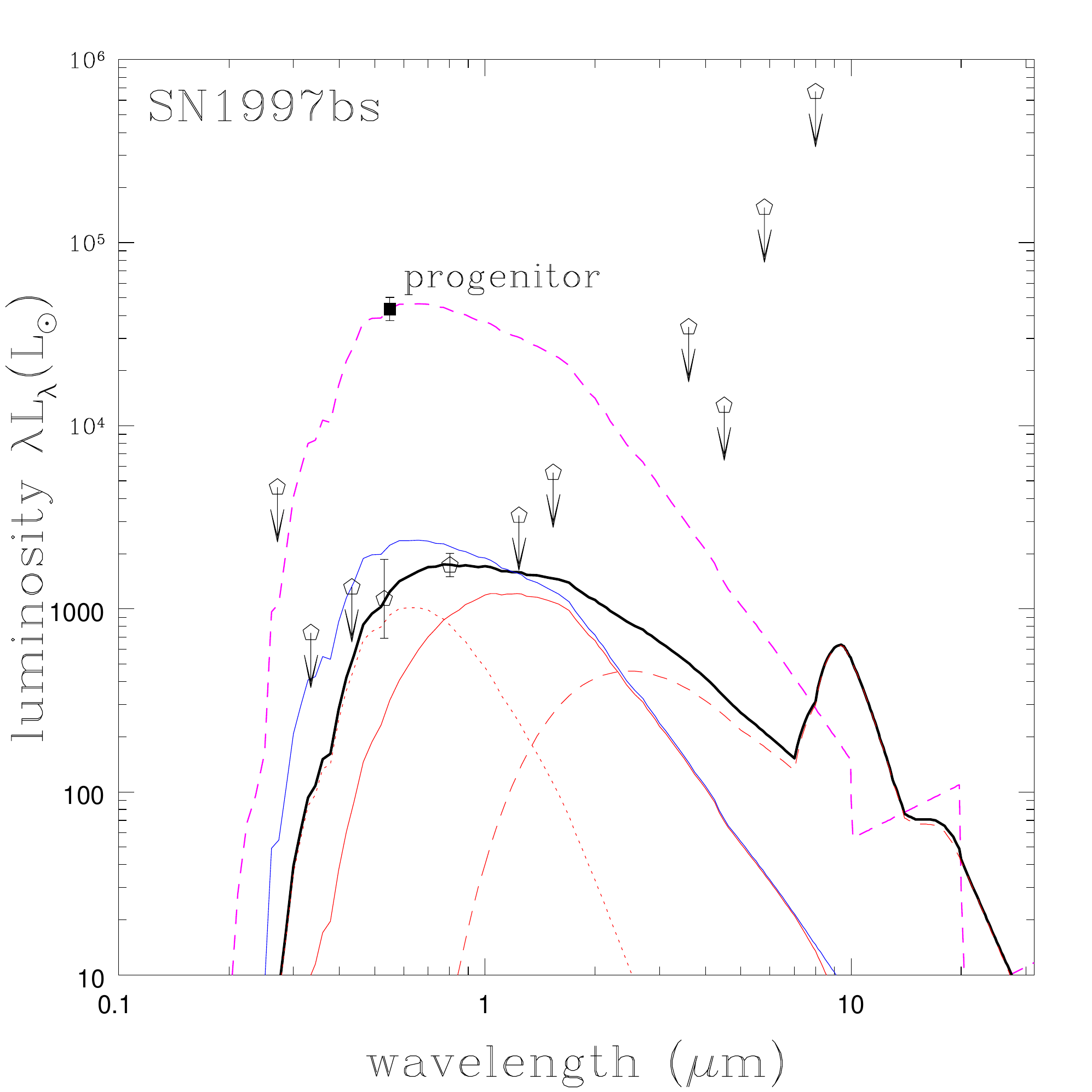}
  \else
    \includegraphics[width=8.6cm, angle=0]{fig5.ps}
  \fi
  \caption{The SED of SN 1997bs with the best-fit model for a silicate wind with an inner edge temperature of $T_{\mathrm{f}} = 1500$ K when treating our $F555W$ and $F814W$ photometry as detections and the other bands as upper limits.  The \emph{HST} and \emph{SST} detections and limits for a surviving star are shown as the black open pentagons with the best-fit observed model spectrum shown as the thick black line, the intrinsic input SED is the thin blue line, and the contributions of the attenuated input radiation, scattered radiation, and dust emission are shown as thin red solid, dotted, and dashed lines respectively.  This particular model has $T_{*}=5252$ K, $\tau_{V,\mathrm{tot}}=1.95$, and $L_{*}=3197~L_{\odot}$.  While the mid-IR limits are relatively weak, the near-IR constraints on hot dust prevent solutions with much higher luminosities and optical depths.  For comparison, we show the \emph{HST} progenitor detection as the solid black square and a model spectrum also with $T_{*}=5252$ K and $L_{*}=10^{4.80}~L_{\odot}$ as the dashed magenta line. \label{fig:best_silicate_wind_sed}}
\end{figure}

Rather than being obscured by a dusty shell ejected during the outburst in 1997, the surviving star could be obscured by a steady-state dusty wind that began following the transient.  For our standard models we adopt an inner (dust formation) temperature of $T_{\mathrm{f}}=1500$ K.  We present, as an example, the best-fitting SED for a wind when treating our $F555W$ and $F814W$ photometry as detections and the other bands as upper limits in Fig. \ref{fig:best_silicate_wind_sed}.  The striking fact is that the luminosity of the best-fitting SED for our candidate detection is fainter than the progenitor by nearly a factor of 20 when modeled as being obscured by a dusty wind.

\begin{figure*}
  \begin{center}
  \ifpdflatex
    \includegraphics[width=0.78\textwidth]{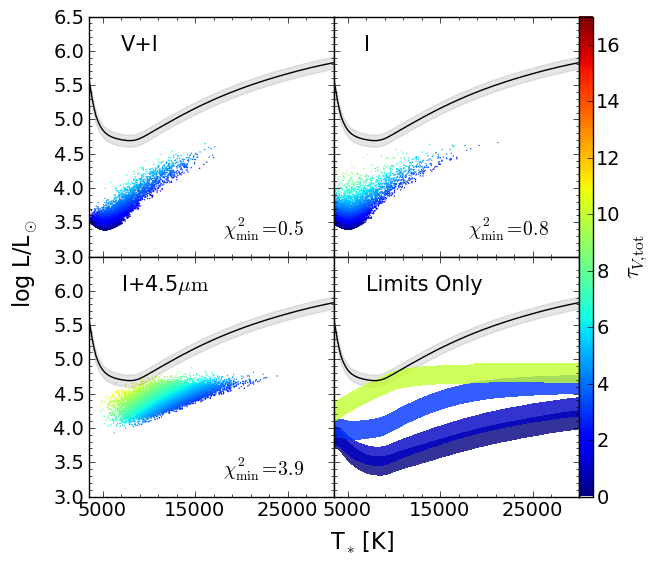}
  \else
    \includegraphics[width=0.78\textwidth]{../../dusty/mylimits/fig6.ps}
  \fi
  \end{center}
  \caption{MCMC results for the luminosity and temperature of a surviving star obscured by a steady-state wind when imposing different possible photometric constraints: treating the \emph{HST} WFC3/UVIS $F555W$ and $F814W$ photometry as detections and the other filters as only upper limits (top left), treating the \emph{HST} WFC3/UVIS $F814W$ photometry as a detection and the other filters as upper limits (top right), treating the \emph{HST} WFC3/UVIS $F814W$ and \emph{SST} $4.5~\mu$m photometry as detections and the other filters as upper limits (bottom left), and treating all photometry as upper limits (bottom right).  
The shaded bands in the `Limits Only' panel show the luminosities within the 90-99.99\% confidence intervals ($6.25 < \Delta \chi^2 < 21.1$ for three parameters -- $L_{*}$, $T_{*}$, and $\tau$ -- relative to no surviving star) for $\tau_{V,\mathrm{tot}}=$ 0, 1, 3, and 10.
For comparison, the solid black line and gray band indicates the progenitor luminosity as constrained by the pre-explosion measurement of $m_{F606}=22.86$ and its $1\sigma$ uncertainty (0.16 mag).  The lowest $\chi^2$ value for the 10,000 MCMC steps accepted is given for each panel.  Well-fitted wind models are possible when treating the $F555W$ and $F814W$ photometry as detections, but these models require a surviving star to be much fainter than the progenitor.  Though the panel where both the $F814W$ and the $4.5~\mu$m photometry are treated as detections appears to allow a luminous survivor, these models have a high $\chi^{2}$ because of the great difficulty in accounting for the $4.5~\mu$m flux with $T_{\mathrm{f}} = 1500$ K dust without violating the \emph{HST} near-IR limits.  It is unlikely that a surviving star is obscured by a steady-state wind unless the star is significantly fainter than the progenitor. \label{fig:wind}}
\end{figure*}

The complete MCMC results are shown in Fig. \ref{fig:wind}.
The \emph{HST} $F110W$ and $F160W$ photometric limits strongly constrain the amount of hot dust around the source.  As a result, at least for the cases where only optical filters are treated as detections, a surviving star is constrained to be cool with low to moderate obscuration ($\tau_{V,\mathrm{tot}} \lesssim 7$) and a significantly lower luminosity than the progenitor.  While the case where the $F814W$ and $4.5~\mu$m photometry are treated as detections appears to favor a heavily obscured source with only slightly lower luminosity than the progenitor, all of the models in this case have high $\chi^{2}$ values because of the great difficulty in accounting for the $4.5~\mu$m flux with $T_{\mathrm{f}} = 1500$ K without violating the \emph{HST} near-IR limits.  When we model all of the photometry as upper limits (see the lower-left panel in Fig. \ref{fig:wind}) we see that $\tau_{V,\mathrm{tot}}\gtrsim10$ is needed to allow the star to survive with an undiminished luminosity behind a dusty wind.  As illustrated by the high $\chi^{2}$ values of the case where the $F814W$ and $4.5~\mu$m photometry are treated as detections, such a high optical depth is only possible if all the optical constraints are only upper limits.

These limits are relatively robust to variations in the assumed dust condensation temperature in the wind.  Models of a steady-state wind with cooler dust condensation temperatures result in slightly weaker luminosity limits for a hot, heavily obscured surviving star.  Even if we drop the condensation temperature to $T_{\mathrm{f}} = 1000$ K (750 K) the luminosity limit for a $T_{*}>7500$ K surviving star with $\tau_{V,\mathrm{tot}}=10$ only increases by $\sim0.3$ dex ($\sim0.5$ dex).  We note, however, that setting $T_{\mathrm{f}} = 1000$ K significantly improves the fit of the model ($\chi_{\mathrm{min}}=0.9$) in the case where $F814W$ and $4.5~\mu$m are treated as detections by easing the tension with the near-IR flux limits.

We can use Eqn. \ref{eqn:mdot} to estimate the mass-loss rate required to achieve $\tau_{V,\mathrm{tot}}\sim10$.  Using the observed ejecta velocity of SN 1997bs as $v_{\mathrm{w}}$, the mass-loss rate would correspond to $3\times 10^{-4}~M_{\odot}\>\mathrm{yr}^{-1}$ ($4\times 10^{-3}~M_{\odot}\>\mathrm{yr}^{-1}$) for a $10^{4.7}~L_{\odot}$ ($10^{5.4}~L_{\odot}$), 7500 K (20,000 K) star.  Such high mass-loss rates are typical of active LBVs \citep{Puls08}, although a still higher mass-loss rate of $\gtrsim 10^{-2.5}~M_{\odot}\>\mathrm{yr}^{-1}$ is needed in order for dust formation to occur around hot ($T\gtrsim 10,000$ K) stars \citep{Kochanek11,Kochanek14}.  Additionally, at these high mass-loss rates, the wind itself becomes optically thick independent of any dust leading to a pseudophotosphere with an apparent temperature of $T \sim 7000$ K \citep{Davidson87}.

A mass loss rate of $\dot{M}>10^{-2.5}~M_{\odot}\>\mathrm{yr}^{-1}$ would correspond to $\tau_{V,\mathrm{tot}}>140$ (12) for $T_{*} = 7500$ K (20,000 K).  Essentially this means that for a hot surviving star there must be either no dust in the wind or a very thick dusty wind ($\tau_{V,\mathrm{tot}}>12$).  
  However, $\tau_{V,\mathrm{tot}}>>10$ is inconsistent with any optical detection of a source.  

\section{Discussion}
\label{sec:conclusions}
The new \emph{HST} images we present clearly lack a source with optical flux comparable to the progenitor of SN 1997bs.  If SN 1997bs was a non-terminal event the diminished optical flux could be due to increased obscuration of the star, an increase in the star's temperature, or an intrinsic decline in the luminosity.  The alternative is that SN 1997bs was a genuine SN.  We evaluate these different scenarios and the limits placed on them by the data.

\subsection{Obscuration}

The late-time photometric evolution of SN 1997bs poses a challenge to
the prevailing notion of it being an SN impostor in which a non-terminal eruption ejected a significant amount of mass that then obscured the surviving star.  The observed limits on the variability at the position of SN 1997bs from \emph{HST} and LBT rule out any expanding dusty shell scenario in which the optical emission represents a `recovery' of the luminous progenitor (see the right-hand panels of Fig. \ref{fig:shell}).  The data cannot, however, rule out the possibility that a surviving star is still cloaked by a dusty shell with $\tau_{V,\mathrm{tot}}>10$, but this scenario is difficult to reconcile with the earlier color evolution of the transient.

The new data are also difficult to reconcile with the alternative, a steady dusty wind, which \cite{Kochanek12} found was consistent with earlier archival data (see Fig. \ref{fig:wind}).
The high mass loss rate needed for self-shielding so that dust can form in the wind of a hot ($T_{*}\gtrsim10,000$ K) star would result in a large $\tau_{V,\mathrm{tot}}$ that would not allow the candidate optical detections reported here and by \cite{Li02} and \cite{vanDyk12}.

\subsection{Bolometric Correction}
If we assume that there is no dust obscuration, the maximum luminosity of the surviving star is $\sim10^{4.2}~L_{\odot}$ (see the $\tau_{V,\mathrm{tot}}=0$ line in the `Limits Only' panel of Fig. \ref{fig:shell} or \ref{fig:wind}), which is significantly fainter than the traditional range of luminosities for LBVs.
It is unlikely that the low luminosity could solely be the result of a large bolometric correction.  Allowing a surviving star to have an effective temperature of up to 40,000 K, the maximum luminosity of a surviving star is still at least a factor of three fainter than the minimum luminosity of the progenitor.
For temperatures higher than 40,000 K, the maximum luminosity of the star can be constrained by the maximum possible H$\beta$ flux allowed by our $F555W$ photometry, since a very luminous hot star would produce a significant flux of ionizing photons and the ejecta would provide an absorbing medium.  If we assume that the star emits a blackbody spectrum and is surrounded by enough gas to absorb the photons with 11\% of recombinations going through the H$\beta$ branch (based on Case B recombination coefficients for $T_{*}=20,000$ K from \cite{Draine11}) and set the H$\beta$ flux equal to our flux limit for $F555W$ (we do not use H$\alpha$ because its wavelength places it in the tail of the $F814W$ filter response curve), we find that the maximum luminosity of a $T_{*}=40,000$ K star is $\sim10^{5.0}~L_{\odot}$, with lower maximum luminosities for increasing $T_{*}$.  Accordingly, we cannot completely rule out the possibility that a progenitor initially with $4400 \lesssim T_{*} \lesssim 13,000$ K survived the episode as a $T_{*}>40,000$ K Wolf--Rayet star.  However, 
we are unaware of any theoretical studies suggesting
that a single short duration event could affect such a dramatic change.

\subsection{A Tuckered-Out Star}
Another possibility that would allow SN 1997bs to be a non-terminal event is that the progenitor survived the event with a diminished luminosity -- the `tuckered-out' star hypothesis \citep{Smith11}.  If a surviving star is unobscured and its temperature is unchanged from pre-eruption, it must be at least a factor of $\sim30$ fainter than the progenitor.  No known mechanism could so drastically diminish the intrinsic luminosity of the star.  As mentioned in the previous section, even if we allow for a significant increase in the temperature of the star, a factor of four decrease in the luminosity is required by the latest observations.  
A conspiracy of factors could explain the dramatic drop in apparent luminosity.
Though an expanding dusty shell is ruled out by limits on the observed variability, a combination of a steady-state dusty wind and a significant increase in $T_{*}$ could allow for a surviving star with a luminosity that has declined by as little as $\sim25\%$.  However, this collusion is disfavored by the candidate late-time detections of an optical source.

The `buildup' or `recovery' time-scale for the radiated energy budget given by \cite{Smith11} is $t_{\mathrm{rad}} = E_{\mathrm{rad}} / L_{*} = t_{1.5} \zeta L_{\mathrm{peak}} / L_{*}$, where $E_{\mathrm{rad}}$ is the energy radiated during the outburst, $L_{\mathrm{peak}}$ is the peak luminosity of the outburst, $L_{*}$ is the quiescent pre-outburst luminosity of the progenitor, and $\zeta$ is a dimensionless factor.  For SN 1997bs this is
\begin{equation}
t_{\mathrm{rad}} \sim 27 \left( \frac{t_{1.5}}{45~\mathrm{days}} \right) \left( \frac{L_{\mathrm{peak}} / L_{*}}{220} \right) \zeta~ \mathrm{yr} ,
\end{equation}
which means that a large fraction of this recovery time-scale has already elapsed with no re-brightening of a surviving star.

This time-scale may not be the most relevant time-scale for the stellar luminosity, since the envelope would likely return to thermal equilibrium primarily through Kelvin--Helmholtz contraction rather than by energy radiated from the core.  In order to decrease the bolometric luminosity for such an extended period of time, either some of the nuclear energy generated must be diverted into gravitational potential energy by inflating the stellar envelope or the nuclear luminosity must have decreased. 

If the missing luminosity is being used to alter the structure of the outer layers of the star, the time-scale for it to lift mass $\Delta M$ against the gravitational potential of the star is
\begin{eqnarray}
t_{\mathrm{infl}} \sim 120 \left( \frac{\Delta M}{M_{\odot}} \right) \left( \frac{M_{*}}{20~M_{\odot}} \right) \left( \frac{T_{*}}{10,000~\mathrm{K}} \right)^{2} \nonumber \\ \times \left( \frac{L_{*}}{10^{5}~L_{\odot}} \right)^{-3/2} \left( \frac{0.5}{f} \right)~\mathrm{yr} ,
\end{eqnarray}
where $M_{*}$, $T_{*}$, and $L_{*}$ are the mass, temperature, and luminosity of the progenitor, and $f$, is the fraction by which the radiated luminosity has decreased.  
This could account for the missing energy, but it is unclear if there is a theoretical mechanism that could drive this.  Moreover, $t_{\mathrm{infl}}$ is not so long that we would expect no external changes on time-scales of a decade.
Another problem is that the most natural post-eruption state would be a star with an overexpanded envelope rather than the reverse.  The envelope would then shrink on a thermal time-scale, rapidly at first but then slowing, making the star overluminous, not subluminous.  The overexpansion is a natural consequence of any transient mechanism which has no `knowledge' of the escape speed and has mainly been discussed in the context of shock-heating non-degenerate companions of Type Ia SNe \citep{Pan13,Shappee13}.

The final alternative is for the nuclear luminosity to have decreased.  The core luminosity should be unaltered as it can only change over a (core) thermal time-scale, but it might be possible for the luminosity from shell-burning to decrease suddenly in certain (fine-tuned) situations.  Perhaps the progenitor had previously experienced enough mass loss for its hydrogen-burning shell to be close to its surface and the star experienced a final shell flash, analogous to the final shell flash of an asymptotic giant branch star.
If the progenitor was very massive star ($M_{i} \gtrsim 80~M_{\odot}$), a thermonuclear outburst from pulsational pair instability might be another possibility \citep{Woosley15}.

\subsection{SN and Confusion}
Alternatively, SN 1997bs could have been a terminal event.
If the star went supernova, our detection of an $F814W = 25.49 \pm 0.12$ source 0$\farcs$016 from the position of SN 1997bs must be due to confusion, a surviving companion, or the SN ejecta.  As we describe in \S\ref{sec:confusion}, the likelihood of our detection being due to confusion is between 4\% and 30\%.  The likelihood of the source being a surviving binary companion is comparable, at 5--20\%, although we would expect a surviving companion to be bluer.
The slight decrease in luminosity between the 2001 \emph{HST}/WFPC2, 2009 \emph{HST}/ACS, and 2013 \emph{HST}/WFC3 $F555W$ and $F814W$ images reduces the likelihood that the detection is due to confusion, although there could be systematic issues in comparing the crowded field photometry based on the three different instruments and the LBT variability limits are consistent with zero.

The $4.5~\mu$m flux measured at the location of SN 1997bs dropped from a significant $31\pm4 \mu$Jy in 2004 to $7\pm3 \mu$Jy in 2014.  This variability indicates that the $4.5~\mu$m flux observed in 2004 originated from SN 1997bs and is not confusion.  The decrease in the $4.5~\mu$m flux while the $3.6~\mu$m flux stayed constant is difficult to explain if the $3.6~\mu$m flux is from SN 1997bs.  More likely, the $3.6~\mu$m flux is unrelated to SN 1997bs (SN 1997bs is close to a spiral arm) and the decreasing $4.5~\mu$m emission was from warm ($\sim 1000$ K) dust that has since cooled.

It is possible that we are detecting residual flux from the shock interaction of SN 1997bs with its circumstellar medium (CSM).  If we parametrize the fraction of the maximum possible shock luminosity radiated in our observed filters as $f$, then the observed shock luminosity is given by
\begin{eqnarray}
L_{\mathrm{s,obs}} \simeq 2.3 \times 10^3 \left( \frac{\dot{M}}{10^{-4}M_{\odot}} \right) \left( \frac{v_{\mathrm{e}}}{765~\mathrm{km\>s^{-1}}} \right)^{3} \nonumber \\ \times \left( \frac{v_{\mathrm{w}}}{100~\mathrm{km\>s^{-1}}} \right)^{-1} \left( \frac{f}{0.1} \right) L_{\odot}
\end{eqnarray}
where $\dot{M}$ is the pre-eruption mass loss rate, $v_{\mathrm{e}}$ is the velocity of the SN ejecta, and $v_{\mathrm{w}}$ is the pre-eruption wind velocity.  Although we do not have constraints on $\dot{M}$ or $v_{\mathrm{w}}$ since little is known about the progenitor, reasonable values of $v_{\mathrm{w}}$ and $f$, together with a relatively (but not outrageously) high $\dot{M}$ during the decades proceeding the transient, are sufficient to possibly account for our detection ($\sim 1700~L_{\odot}$ in $F814W$).

It is worth reconsidering why SN 1997bs was originally designated as a likely SN impostor: a low peak luminosity and a possible flattening of the late-time ($\gtrsim 250$ days post-explosion) light curve at $\sim 0.5$ mag fainter than the progenitor \citep[see][]{vanDyk99}.  Though the source continued to fade well beyond this level, \cite{vanDyk12} re-asserted the case for a surviving star based on the 2009 \emph{HST} observations in which the star was `recovered' with a brightness consistent with measurements made in 2001 \citep[see][]{vanDyk12}.  However, the source at the coordinates of SN 1997bs, rather than re-brightening as required if the optical emission was from a surviving star partially obscured by an expanding dusty shell, has only continued to fade and is now $\sim4$ mag fainter than the progenitor.  
For the most likely physical conditions, the near-IR limits rule out the detection of optical emission from a star surviving in a dusty wind.
Rather than providing evidence that 97bs was a non-terminal eruption, the late-time observations are difficult to reconcile with this prevailing notion.

SN 1997bs can be explained as a low-energy SN.  Over the last two decades, a class of faint Type IIP SNe has emerged with peak magnitudes as faint as $M_{V}=-13.76$ \citep[SN 1999br;][]{Pastorello04}, less than a magnitude brighter than 97bs ($M_{V}\sim-12.9$).  These faint SNe have low ejected $^{56}$Ni masses ($\sim10^{-3} M_{\odot}$) and require low explosion energies ($L<10^{51}$ erg).  The low $^{56}$Ni mass is likely due to fall-back of material on to the collapsed remnant \citep[see e.g.,][]{Woosley95}.  Although the late-time light curve of SN 1997bs might have significant contributions from a CSM--SN shock interaction, we can estimate the maximum $^{56}$Ni mass ejected by scaling the \emph{V}-band luminosity to that of the well-studied faint IIP SN 1997D at the same epoch (270 days post-explosion).  Nominally this would suggest that that the maximum $^{56}$Ni mass for 97bs is 20 times smaller than the 0.002 $M_{\odot}$ estimated for SN 1997D by \cite{Turatto98}.  However, 97bs was much redder than 97D by this point ($V-I\sim4.5$ versus $\sim1.2$ mag) and was likely significantly obscured.  If we posit that extinction is needed to match the colors, then 97bs had $E(B-V)\sim2.3$ mag at this point, which would increase the maximum $^{56}$Ni mass by nearly a factor of 1000.  Even if the shock luminosity contributed a large fraction of the luminosity, 97bs could still potentially have more ejected $^{56}$Ni than a significant fraction of faint IIP's.

Thus, SN 1997bs could be a IIn analog of a faint IIP.
Progenitor detections and hydrodynamical modeling of light curves point to faint IIP SNe progenitor masses of 10-15 $M_{\odot}$ \citep{Spiro14}.  Meanwhile, recent work in SN theory by \cite{Ugliano12} and \cite{Pejcha15} suggests that stars with masses between 14 and 16 $M_{\odot}$, some between 17 and 22 $M_{\odot}$, and stars between 23 and 26 $M_{\odot}$ are the most difficult to explode.
Faint IIP SNe might arise from the lower mass (14--16 $M_{\odot}$) stars that have density structures resulting in low-energy explosions while 97bs and similar `impostors' could arise from the higher mass (23--26 $M_{\odot}$) stars.  These higher mass stars have the potential to have the pre-explosion mass loss required for a transient to have the narrow hydrogen emission lines that are characteristic of SN IIn (and SN impostor) spectra.

The event could even be a collapse-induced thermonuclear SN \citep{Kushnir14}, which would dump enough energy in the stellar envelope quickly enough to drive a shock in a manner similar to the piston simulations of \cite{Dessart10}.  This would be a weak, terminal event of a star in one of the harder-to-explode mass ranges.

All of these low-energy explosions are also much more favorable for dust formation because condensation will occur at much higher densities \citep[see, e.g.,][]{Kochanek14d}.  This would provide a natural explanation of the increasingly red colors observed by \cite{vanDyk00}, since dust formation fairly naturally correlates with dropping luminosity.  The lack of a surviving star then avoids the problems with dusty models of a surviving star.

\subsection{Summary}
Our late-time observations of SN 1997bs seem most easily understood if SN 1997bs was, in fact, the first SN found by LOSS.  
It is difficult to explain the data as dust obscuration by a wind and it is only possible to explain by an expanding shell if the star is still completely obscured in the optical and the putative recovery of a surviving star is actually unrelated emission.
Similarly, invoking bolometric corrections or simply making the star less luminous seems difficult to support.  While the observations do not rule out a surviving star, they do undermine the evidence in favor of this scenario.  Occam's razor, if nothing else, argues for simply making it an SN rather invoking a more baroque model.

Nature does not, of course, have to respect Occam's razor.  Further observations are necessary to confirm the fate of SN 1997bs.  Unfortunately it is impossible to resolve any IR emission from SN 1997bs with the diffuse emission from the nearby spiral arm with \emph{SST}.  The improved resolution and sensitivity of the \emph{James Webb Space Telescope} are needed to reduce the mid-IR limits enough to completely rule out a cool star behind a thick, dusty wind or shell.

As argued by \cite{Kochanek12}, the current class of SN impostors is likely comprised of multiple phenomena.  We find that SN 1997bs, an archetypal SN impostor, was likely a genuine SN, but there are also several SN impostors that have clearly undergone multiple non-terminal outbursts (e.g., SN 2000ch, 2002kg, 2009ip).  Moreover, in newly obtained \emph{HST} images, we confirm a surviving star to the SN impostor 1954J, as proposed by \cite{Smith01} and \cite{vanDyk05} (Adams \& Kochanek, in preparation).  At least a significant fraction of SN impostor progenitors are relatively low mass (15 $M_{\odot} < M < 25 M_{\odot}$) stars \citep{Smith11,Kochanek12}.  These masses are lower than the mass range of `classical' LBVs and should not be close to their Eddington limits and any of the radiative instabilities of classical LBVs.  Recurrent impostors have the highest progenitor masses (see Table \ref{tab:masses}).  Perhaps some of the lower mass impostors are faint IIn SNe and only the impostors arising from more massive stars are non-terminal eruptions of LBVs.

\ifapj
  \begin{acknowledgements}
\else
  \section*{Acknowledgements}
\fi
We thank Jill Gerke for providing the LBT variability limits, Andrew Dolphin for assistance with {\sc dolphot}, and Rubab Khan, Marc Pinsonneault, Jos\'{e} Prieto, Kris Stanek, and Todd Thompson for discussions and comments.
This work is based in part on observations made with the \emph{Spitzer Space Telescope}, which is operated by the Jet Propulsion Laboratory, California Institute of Technology under a contract with NASA, and in part on observations made with the NASA/ESA \emph{Hubble Space Telescope} obtained at the Space Telescope Institute, which is operated by the Association of Universities for Research in Astronomy, Inc., under NASA contract NAS 5-26555. These observations are associated with program GO-13477.
This work is also based in part on observations made with the Large
Binocular Telescope. The LBT is an international collaboration
among institutions in the United States, Italy, and Germany. The
LBT Corporation partners are: the University of Arizona on behalf
of the Arizona university system; the Istituto Nazionale di
Astrofisica, Italy; the LBT Beteiligungsgesellschaft, Germany,
representing the Max Planck Society, theAstrophysical Institute
Potsdam, and Heidelberg University; the Ohio State University;
and the Research Corporation, on behalf of the University of
Notre Dame, the University of Minnesota, and the University of
Virginia.
\ifapj
  \end{acknowledgements}
\fi

\begin{appendix}
\section{Dust Size Distribution}
\label{app:one}
\input{tableA1.tex}

\begin{figure*}
  \begin{center}
  \ifpdflatex
    \includegraphics[width=0.8\textwidth]{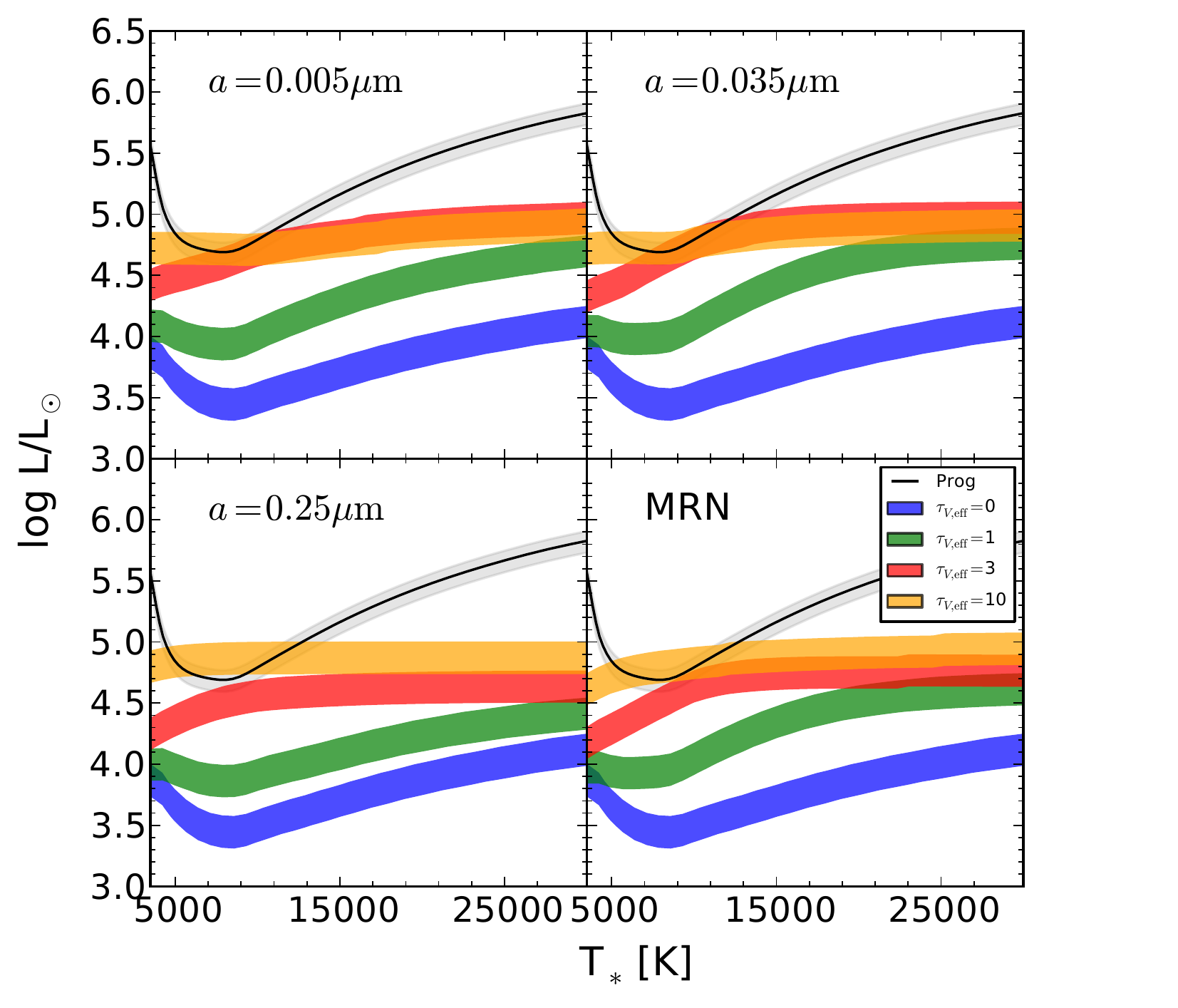}
  \else
    \includegraphics[width=0.8\textwidth]{../../dusty/mylimits/figA1_eff.ps}
  \fi
  \end{center}
  \caption{The luminosities within the 90-99.99\% confidence intervals ($6.25 < \Delta \chi^2 < 21.1$ for three parameters -- $L_{*}$, $T_{*}$, and $\tau$ -- relative to no surviving star) for $\tau_{V,\mathrm{tot}}=$ 0, 1, 3, and 10 with a steady-state silicate wind for different dust grain sizes, $a$, as compared to the results for an MRN dust size distribution (bottom-right panel) when treating all photometry as upper limits.  The solid black line and gray band indicates the progenitor luminosity as constrained by the pre-explosion measurement of $m_{F606}=22.86$ and its $1\sigma$ uncertainty (0.16 mag).  For most temperatures and effective optical depths, using a different (reasonable) dust grain size distribution would only shift the luminosity limit by $\sim0.1$ dex. \label{fig:dustsize}}
\end{figure*}

Although the MRN dust size distribution we assumed for the analysis presented in this paper is a reasonably accurate prescription for the size distribution of interstellar dust, dust formed around a particular star will grow to a particular size that is dependent on the density and radiation field of the circumstellar environment \citep[see, e.g.,][for a discussion]{Kochanek14}.  Unfortunately, we are unable to determine a most likely dust size, since the size, an input to our models, is dependent on the output (i.e., $\tau$, $T_{*}$, and $L_{*}$) from the models.

Instead, we evaluated our models for a number of dust sizes ranging from $0.005~\mu\mathrm{m}$ to $0.25~\mu\mathrm{m}$ and verified that our conclusions are robust to the chosen dust size.  For an example, in Fig. \ref{fig:dustsize} we present the results of the silicate wind model for a range of dust sizes and effective optical depths.  For silicate dust grain sizes, $a$, of $0.005~\mu\mathrm{m}$, $0.035~\mu\mathrm{m}$, $0.25~\mu\mathrm{m}$, and an MRN size distribution the albedo is $w_{V} \simeq$ 0, 0.10, 0.43, and 0.39, corresponding to $\tau_{V,\mathrm{eff}} \simeq$ 1.00, 0.82, 0.32, and 0.37, respectively.  For a given effective optical depth the luminosity limits on a surviving star are largely unchanged.  The most significant difference is that the $\chi^{2}$ of the best-fitting model for a silicate wind with our $F814$ and $4.5~\mu\mathrm{m}$ photometry treated as detections for a dust size of $0.035~\mu\mathrm{m}$ is only 1.3 (compared to 3.9 for an MRN distribution of dust), slightly reducing the confidence with which we can rule out a luminous star surviving behind a dusty wind.

The impacts of the dust size on the results of the shell model are even more limited.  For a given effective optical depth, the only significant change for different dust sizes is that the luminosity limit for a cool star with heavy obscuration is increased.  The key point, however, is that the light curve constraints continue to rule out the optical emission being a detection of a luminous surviving star dimmed by an expanding shell.

\end{appendix}

\bibliography{references}
\ifapj
  \bibliographystyle{apj}
\else
  \bibliographystyle{mn2e}
\fi

\clearpage
\end{document}

%% file: table1.tex
\ifapj
  \begin{deluxetable*}{llllrrrrrrr}
  \tablecaption{{\sc dolphot} Parameters}
  \tablehead{
  \colhead{Description} &
  \colhead{Parameter} &
  \colhead{WFC3/UVIS} &
  \colhead{WFC3/IR}
  }
  \startdata
\else
  \begin{table*}
  \begin{minipage}{10cm}
  \caption{DOLPHOT Parameters}
  \begin{tabular}{llllrrrrrrr}
  \hline
  \hline
  {Description} & {Parameter} & {WFC3/UVIS} & {WFC3/IR} \\
  \hline
\fi
Photometry aperture radius 			& \tt{RAper}		&	4	&	3	\\
Inner sky radius 				& \tt{RSky0}		&	15	&	8	\\
Outer sky radius 				& \tt{RSky1}		&	35	& 	20	\\
$\chi$-statistic aperture size 			& \tt{RChi}		&	2	&	1.5	\\ 
Sky Fit 					& \tt{FitSky}		&	1	&	1	\\
Spacing for sky measurement			& \tt{SkipSky}		&	2	&	2	\\
\ifapj
  \enddata
  \tablecomments{Only parameters differing from those used by \cite{Dalcanton12} are listed here.}
  \label{tab:dolphot}
  \end{deluxetable*}
\else
  \hline
  \hline
  \end{tabular}
  Only parameters differing from those used by \cite{Dalcanton12} are listed here.
  \label{tab:dolphot}
  \end{minipage}
  \end{table*}
\fi

%% file: tableA1.tex
\ifapj
  \begin{deluxetable}{ll}[t]
  \tablecaption{SN Impostor Masses}
  \tablehead{
  \colhead{Object} &
  \colhead{$M_{i} [M_{\odot}]$}
  }
  \startdata
\multicolumn{2}{c}{\textbf{Single Episode}}				\\
SN 2008S        & $\sim10$\tablenotemark{a}				\\
2008-OT         & $12-15$\tablenotemark{a}				\\
SN 1997bs       & $12-24$\tablenotemark{a}, $>20$\tablenotemark{b}	\\
SN 2003hm       & $14-17$\tablenotemark{a}, $>20$\tablenotemark{b}	\\
SN 1954J        & $14-21$\tablenotemark{a}, $>20$\tablenotemark{b}      \\
2009-OT         & $\sim25$\tablenotemark{b}				\\
HD 5980		& $58-79$\tablenotemark{c}\tablenotemark{*}		\\
SN 1961V        & $>80$\tablenotemark{d}				\\
\multicolumn{2}{c}{\textbf{Reccurent}}					\\
P Cygni		& $20-40$\tablenotemark{e}\tablenotemark{*}		\\	
SN 2002kg	& $>15$\tablenotemark{a}, $>40$\tablenotemark{b}	\\ 
SN 2000ch       & $20-80$\tablenotemark{a}, $>40$\tablenotemark{f}	\\
SN 2009ip       & $50-80$\tablenotemark{b}				\\
$\eta$ Car	& $60-120$\tablenotemark{a}				\\
\enddata
\tablenotetext{*}{Current mass}
\tablenotetext{a}{\cite{Kochanek12}}
\tablenotetext{b}{\cite{vanDyk12}}
\tablenotetext{c}{\cite{Foellmi08}}
\tablenotetext{d}{\cite{Kochanek11c}}
\tablenotetext{e}{\cite{Lamers83}}
\tablenotetext{f}{\cite{Wagner04}}
\label{tab:masses}
\end{deluxetable}

\else
  \begin{table}
  \centering{
  \begin{minipage}{5cm}
  \caption{SN Impostor Masses}
  \begin{tabular}{ll}
  \hline
  \hline
  {Object} & {$M_{i} [M_{\odot}]$} \\
  \hline
\multicolumn{2}{c}{\textbf{Single Episode}}                             \\
SN 2008S        & $\sim10^{a}$                             \\
2008-OT         & $12-15^{a}$                              \\
SN 1997bs       & $12-24^{a}$, $>20^{b}$      \\
SN 2003hm       & $14-17^{a}$, $>20^{b}$      \\
SN 1954J        & $14-21^{a}$, $>20^{b}$      \\
2009-OT         & $\sim25^{b}$                             \\
HD 5980         & $58-79^{c*}$             \\
SN 1961V        & $>80^{d}$                                \\
\multicolumn{2}{c}{\textbf{Reccurent}}                                  \\
P Cygni         & $20-40^{e*}$             \\      
SN 2002kg       & $>15^{a}$, $>40^{b}$        \\
SN 2000ch       & $20-80^{a}$, $>40^{f}$      \\
SN 2009ip       & $50-80^{b}$                              \\
$\eta$ Car      & $\sim160^{g}$                             \\
\hline
\hline
\end{tabular}
$^{*}$Current mass \\
$^{a}$\cite{Kochanek12} \\
$^{b}$\cite{vanDyk12} \\
$^{c}$\cite{Foellmi08} \\
$^{d}$\cite{Kochanek11c} \\
$^{e}$\cite{Lamers83} \\
$^{f}$\cite{Wagner04} \\
$^{g}$\cite{Davidson97}
\label{tab:masses}
\end{minipage}
} 
\end{table}
\fi